\documentclass[english,pra,manuscript]{revtex4}
\usepackage[T1]{fontenc}
\usepackage[latin9]{inputenc}
\usepackage{amsmath}
\usepackage{amssymb}

\makeatletter
\@ifundefined{textcolor}{}
{%
 \definecolor{BLACK}{gray}{0}
 \definecolor{WHITE}{gray}{1}
 \definecolor{RED}{rgb}{1,0,0}
 \definecolor{GREEN}{rgb}{0,1,0}
 \definecolor{BLUE}{rgb}{0,0,1}
 \definecolor{CYAN}{cmyk}{1,0,0,0}
 \definecolor{MAGENTA}{cmyk}{0,1,0,0}
 \definecolor{YELLOW}{cmyk}{0,0,1,0}
 }

\makeatother

\newtheorem{thm}{Theorem}   \newtheorem{lem}{Lemma}

\makeatother

\usepackage{babel}

\makeatother

\usepackage{babel}

\begin{document}

\title{Entanglement and perfect LOCC discrimination of a class of multiqubit
states }

\author{Somshubhro Bandyopadhyay}

\affiliation{DIRO, Université de Montréal, C.\,P.~6128, Succursale Centre-Ville,
Montréal, Québec, Canada H3C 3J7}

\altaffiliation{Present address: Department of Physics, Bose Institute, Kolkata 700009, India}

\email{som@bosemain.boseinst.ac.in}
\begin{abstract}
It is shown that while entanglement ensures difficulty in discriminating
a set of mutually orthogonal states perfectly by local operations
and classical communication (LOCC), entanglement content does not.
In particular, for a class of entangled multi-qubit states, the maximum
number of perfectly LOCC distinguishable orthogonal states is shown
to be independent of the average entanglement of the states, and the
spatial configuration with respect to which LOCC operations may be
carried out. It is also pointed out that for this class, the make-up
of an ensemble, that is, whether it consists only of entangled states
or is a mix of both entangled and product states, determines the maximum
number of perfectly distinguishable states. 
\end{abstract}
\maketitle
Suppose a multipartite quantum state, secretly chosen from a set of
mutually orthogonal states, is distributed amongst several observers
who are given the task to identify the state without making any error.
If the observers are located in the same laboratory, they can perform
collective measurements to correctly identify the given state with
certainty. However, within the framework of local operations and classical
communication (LOCC), wherein they can only perform arbitrary quantum
operations on their respective subsystems and communicate by classical
channels but are not allowed to exchange quantum states, they may
not be able to perfectly distinguish even mutually orthogonal vectors.
This shows, at a very basic level, the limitation of LOCC to extract
the entire quantum information encoded in a global quantum state.
To what extent this global information can be reliably accessed locally
then essentially boils down to the problem of faithful discrimination
of mutually orthogonal vectors by LOCC \cite{WH2000,Ghosh2001,Ghosh2002,Horodecki2003,Fan,BW2006}.

A fundamental result is due to Walgate \emph{et al} who proved that
any set of two orthogonal quantum states can always be perfectly discriminated
by LOCC \cite{WH2000} irrespective of their entanglement and multipartite
structure. However, for sets containing more than two mutually orthogonal
states, perfect distinguishability is not always possible and examples
can be found in both orthogonal product \cite{Bennett99,Bennett-UPB}
and entangled ensembles \cite{Ghosh2001,Ghosh2002,Horodecki2003,Fan}.
It is important to note that only a set of entangled states can be
completely indistinguishable, that is, it is not possible to correctly
identify even one state with a non-vanishing probability whereas a
set containing at least one product state is always conclusively distinguishable
\cite{Chefles2003,BW2006}, that is, the set contains at least one
state (trivially the product state, possibly more) that can be correctly
identified with a non-zero probability. Examples of completely indistinguishable
sets include the two qubit Bell basis, and more generally, any entangled
bipartite basis.

One of the main focuses of LOCC distinguishability of quantum states,
regardless of their bipartite or multipartite structure, is to understand
the extent to which entanglement is responsible for their indistinguishability.
The evidence so far, as noted above, is, at best, mixed although intuitively
we expect weakly entangled states should be somewhat more distinguishable
than those strongly entangled in the sense weakly entangled states
are {}``less'' non-local. A step towards quantifying the relation
between entanglement and LOCC distinguishability was taken in a recent
work by Hayashi \emph{et al} \cite{Hayashi2006} who observed that
the number of pure states that can be perfectly discriminated by LOCC
is bounded above by the total dimension over average entanglement.
In particular, if the set of states $\{|\phi_{i}\rangle\}_{i=1}^{N}$
is perfectly distinguishable by LOCC, then the average $\overline{E(|\phi_{i}\rangle)}$
of entanglement {}``distances'' $E(|\phi_{i}\rangle)$ must be less
than the total dimension $D/N$, that is, $N\leq D/\overline{E(|\phi_{i}\rangle}$,
where {}``entanglement distances'' are appropriately defined in
terms of an entanglement measure. In its exact form the inequality
is hierarchial with respect to different measures of entanglement,
\begin{equation}
N\leq\frac{D}{\overline{1+R(|\phi_{i}\rangle)}}\leq\frac{D}{\overline{2^{E_{R}(|\phi_{i}\rangle)}}}\leq\frac{D}{\overline{2^{E_{g}(|\phi_{i}\rangle)}}}\label{eq:Hayashi}\end{equation}
 where, corresponding to the state $|\phi_{i}\rangle$, $R(|\phi_{i}\rangle)$
is the global robustness of entanglement \cite{robust}, $E_{R}(|\phi_{i}\rangle)$
is the relative entropy of entanglement \cite{R-entropy}, $E_{g}(|\phi_{i}\rangle)$
is an extension of the geometric measure of entanglement \cite{G-Measure},
and $\overline{x_{i}}=\frac{1}{N}\sum_{i=1}^{N}x_{i}$ denotes the
average. 

The above inequality shows that it is not possible to perfectly distinguish
an arbitrary number of mutually orthogonal entangled states. If a
set of entangled states is perfectly distinguishable by LOCC then
the cardinality of the set must not violate the above inequality.
Perhaps more importantly the upper bound which is inversely proportional
to the average entanglement of the ensemble indicates the possibility
to perfectly distinguish a larger number of weakly entangled states
than strongly entangled states. Here one should note that the inequality
stretches the upper bound only up to the total dimension as average
entanglement approaches zero, as one would expect, but says nothing
whether the upper bound is achieved.

In a multipartite setting, besides entanglement, there is another
crucial component, viz. the spatial configuration pertaining to a
given LOCC protocol, that plays a significant role in distinguishability/indistinguishability
of a set of states. Generally speaking, the setting where every party
is separated from one another imposes maximum constraint, thereby
causing states to be more indistinguishable, whereas in other configurations,
most notably those that allow collective operations (like bipartitions),
states tend to be more distinguishable. As a result, typically, states
that were not perfectly distinguishable before, become perfectly distinguishable
when collective operations are allowed. For example, the following
three qubit mutually orthogonal states, $\frac{1}{\sqrt{2}}|000\rangle_{ABC}\pm|111\rangle_{ABC},\frac{1}{\sqrt{2}}|011\rangle_{ABC}\pm|100\rangle_{ABC}$,
are completely indistinguishable in the A-B-C and A-(BC) formations,
but are perfectly distinguishable across B-(AC) and C-(AB) bipartitions.
These complexities, which are absent in a bipartite situation, greatly
enhances the difficulty to obtain non-trivial bounds on the maximum
number of perfectly distinguishable states that hold for all spatial
configurations. On the other hand, precisely these properties can
be cleverly exploited to devise multipartite cryptography primitives
like secret sharing \cite{Markham} and data hiding, and therefore
it is very much desirable to search for robust upper bounds, even
if they are only applicable for a class of states.

The set of mutually orthogonal multipartite states that we consider
in this paper are the canonical $N-$qubit GHZ states. The complete
basis can be represented as a collection of $2^{N-1}$ conjugate pairs
in which the $i^{th}$ conjugate pair is given by, \begin{eqnarray}
|\psi_{i}^{+}\rangle & = & \alpha_{i}|\mathbf{k_{i}}\rangle+\beta_{i}|\mathbf{\overline{k_{i}}}\rangle\nonumber \\
|\psi_{i}^{-}\rangle & = & \beta_{i}|\mathbf{k_{i}}\rangle-\alpha_{i}|\mathbf{\overline{k_{i}}}\rangle,i=1,...,2^{N-1}\label{eq:ith-conjugate-pair}\end{eqnarray}
 where, $\alpha_{i}\geq\beta_{i}$ are real, and satisfy the normalization
$\alpha_{i}^{2}+\beta_{i}^{2}=1,\,\forall i$, $\mathbf{k}$ is a
N-bit string of $\{0,1\}$ and $\mathbf{\overline{k}}$ is its bitwise
orthogonal. We show that for the above class of states, entanglement
content is not a key factor, as one would expect from inequality (\ref{Hayashi-II}),
in determining the maximum number of perfectly distinguishable states
by LOCC. We find that the maximum number of perfectly distinguishable
states by LOCC is $2^{N-1}$ and this does not depend on the average
entanglement of the states as long as entanglement is non-zero for
every state. This is to say, neither an ensemble of maximally entangled
GHZ states nor an ensemble of GHZ states having vanishingly small
entanglement would allow more than $2^{N-1}$ states to be perfectly
discriminated by LOCC. Moreover this threshold value is maximally
robust, in the sense, it holds for all conceivable spatial configurations
including every bipartition.

We further show that it is the make up of the ensemble that decides
the maximum number of perfectly distinguishable states. That is, the
threshold value gradually approaches the total dimension as more and
more product states are included in the set. Therefore, if we insist
less entanglement leads to more perfectly distinguishable states by
LOCC, then it may be achieved, as it does in our example, only by
sacrificing entanglement of some states altogether, and not by reducing
entanglement of every state to a vanishingly small amount.

To compute entanglement, we choose relative entropy \cite{R-entropy},
which, for the above set of states can be exactly obtained and is
given by the entanglement entropy \cite{Ent-Entropy}. For the $i^{th}$
conjugate pair, entanglement is simply, $E_{i}^{\pm}=-\alpha_{i}^{2}\log_{2}\alpha_{i}^{2}-\beta_{i}^{2}\log_{2}\beta_{i}^{2}$
and for any subset of states $S$ of cardinality $|S|$, the average
entanglement $\overline{E_{S}}=\frac{1}{|S|}\sum_{i,sign}E_{i}^{sign}$
can be smoothly varied between 0 and 1 ebit. For our choice of the
measure of entanglement, the relevant inequality that sets an upper
bound on the number of perfectly distinguishable states by LOCC is
simply, \begin{equation}
N\leq\frac{D}{\overline{2^{E(|\phi_{i}\rangle)}}}\label{Hayashi-II}\end{equation}

For any pure state $|\phi_{i}\rangle$ chosen from the N-qubit GHZ
ensemble, $E(|\phi_{i}\rangle)$ lies between $0$and $1$. Thus,
$2^{E(|\phi_{i}\rangle)}$ lies between $1$ and $2$, and therefore
$\overline{2^{E(|\phi_{i}\rangle)}}$ must also lie between $1$ and
2. For maximally entangled GHZ states $(\alpha_{i}=\beta_{i}=1/\sqrt{2}\:\forall i)$
for which the entanglement of every state is simply 1 ebit, it is
easy to see that the maximum number of perfectly LOCC distinguishable
maximally entangled GHZ states (or their local unitary equivalents)
is $2^{N-1}$ (the bound is tight \cite{Hayashi2006}), whereas, for
the computational basis ($\beta_{i}=0\,\forall i$), the inequality
suggests an upper bound equal to the total dimension when average
entanglement is zero, and indeed, the entire basis is perfectly distinguishable
by LOCC. Our goal is to find out the maximum number of perfectly distinguishable
GHZ states in the regime $0<\overline{E_{S}}\leq1$. We first consider
all-entangled ensembles where $E_{i}^{\pm}>0\:\forall i$.

\begin{thm} Let $S$ be any set of states chosen from the $N$ qubit
GHZ basis given by Eq. (\ref{eq:ith-conjugate-pair}), such that $E_{i}\neq0,\forall i$.
If there is a spatially separated configuration, where the set is
perfectly LOCC distinguishable, then $|S|\leq2^{N-1}$. This upper
bound is tight. \end{thm}

Although, the validity of the upper bound for all sets of entangled
GHZ states requires that every state in the ensemble be entangled,
the upper bound, itself, is independent of the average entanglement
$\overline{E_{S}}$, where $0<\overline{E_{S}}\leq1$. Let us also
emphasize that the upper bound holds across all bipartitions, and
therefore for any spatially separated configuration. The upper bound
is further shown to be tight by showing the existence of an unique
set $S$, $|S|=2^{N-1}$ of states, comprising one state from each
conjugate pair. This set is perfectly distinguishable by LOCC under
the least favorable configuration, namely when every qubit is separated
from each other. Remarkably once the threshold value is exceeded,
the set ceases to be perfectly distinguishable even when collective
operations are allowed except when all qubits are together.

It follows from a result in Ref. \cite{Horodecki2003} that every
bipartite entangled basis is completely indistinguishable. Therefore,
the entangled GHZ basis is also completely indistinguishable in every
spatial configuration as it is completely indistinguishable across
every bipartition. It would be interesting to know if there is a non-trivial
threshold beyond which any ensemble of GHZ states is completely indistinguishable
for a particular spatial configuration, if not for all. The following
result, however, negates that possibility.

\begin{thm} There always exist conclusively distinguishable sets
$S$, $2^{N-1}\leq|S|<2^{N}$. Moreover such sets are conclusively
distinguishable even when all qubits are spatially separated.\end{thm}

Our final result shows that the make up of an ensemble is critical
in determining the maximum number of LOCC distinguishable GHZ states.
When product states are included in the set, the upper bound approaches
the total dimension. Consider a hybrid basis consisting of $K,K<2^{N-1}$
conjugate pairs- Eq.$\:$(\ref{eq:ith-conjugate-pair}), and $2^{N}-2K$
product states obtained by assigning $\beta_{i}=0$ for some values
of $i$. It is obvious that for any set of states $S,0\leq\overline{E_{S}}\leq1$.

\begin{thm}Let $S$ be any set of states chosen from such a $N$
qubit hybrid basis. If there is a spatially separated configuration
where the set is perfectly LOCC distinguishable, then $|S|\leq2^{N}-K$.
This upper bound is tight. \end{thm}

Therefore, the only way to decrease average entanglement and increase
the number of perfectly distinguishable states at the same time is
to get rid of entanglement of some states altogether. For an all-entangled
ensemble the upper bound is always $2^{N-1}$, no matter how small
or large the average entanglement is. On the other hand, if the ensemble
is hybrid, one can perfectly distinguish more states even though its
average entanglement could be considerably higher than all-entangled
ensembles.

We now prove our results. We begin with two useful lemmas.

\begin{lem} The following set of four mutually orthogonal normalized
two-qubit states \begin{eqnarray}
|\psi_{1}^{+}\rangle & = & \alpha_{1}|00\rangle+\beta_{1}|11\rangle;\alpha_{1}\geq\beta_{1}>0\nonumber \\
|\psi_{1}^{-}\rangle & = & \beta_{1}|00\rangle-\alpha_{1}|11\rangle\nonumber \\
|\psi_{2}^{+}\rangle & = & \alpha_{2}|01\rangle+\beta_{2}|10\rangle;\alpha_{2}\geq\beta_{2}>0\label{eq:two-qubit}\\
|\psi_{2}^{-}\rangle & = & \beta_{2}|01\rangle-\alpha_{1}|10\rangle\nonumber \end{eqnarray}
 is completely indistinguishable, and any subset of three states is
not perfectly distinguishable.\end{lem}

The first part of the lemma was proved in \cite{Ghosh2002,Horodecki2003}.
The proof of the second part follows from \cite{Ghosh2002,BW2006}.

\begin{lem}The following set of four mutually orthogonal two qubit
states \begin{eqnarray}
|\psi^{+}\rangle & = & \alpha_{1}|00\rangle+\beta_{1}|11\rangle;\alpha_{1}\geq\beta_{1}>0\nonumber \\
|\psi^{-}\rangle & = & \beta_{1}|00\rangle-\alpha_{1}|11\rangle\nonumber \\
|\phi\rangle & = & |01\rangle\label{eq:two-qubit-mixed}\\
|\chi\rangle & = & |10\rangle\nonumber \end{eqnarray}
 and its following subsets , $\{|\psi^{\pm}\rangle,|\phi\rangle\}$,
$\{|\psi^{\pm}\rangle,|\chi\rangle\}$ are not perfectly distinguishable
by LOCC. \end{lem}

For the proof of lemma 2, see \cite{Ghosh2002}.

\emph{Proof of Theorem 1.} To prove that a set of states is not perfectly
distinguishable for all spatial configurations, we must show that
the states are not perfectly distinguishable across any bipartition.
We first prove our result choosing an arbitrary bipartition. Then
we will show how the proof can be worked out similarly for any other
bipartition.

Denote a bipartition $Alice-Bob$ as $(m,Q)$ where $m$ qubits belong
to Alice, and the rest to Bob; $1\leq m\leq N/2$ for even $N$ and
$1\leq m\leq(N-1)/2$ for odd $N$; the index $Q$ represents the
specific set of $m$ qubits that belong to Alice (note that there
are ${N \choose m}$ ways to choose the specific $m$ qubits). Rewrite
a conjugate pair, say, $|\psi_{i}^{\pm}\rangle$ (Eq. (\ref{eq:ith-conjugate-pair})),
to explicitly reflect the bipartite form:

\begin{eqnarray}
|\psi_{i}^{+}(m,Q)\rangle_{AB} & = & \alpha_{i}|\mathbf{m}\rangle_{A}|\mathbf{(N-m)}\rangle_{B}+\beta_{i}|\mathbf{\overline{m}}\rangle_{A}|\mathbf{\overline{(N-m)}}\rangle_{B}\nonumber \\
|\psi_{i}^{-}(m,Q)\rangle_{AB} & = & \beta_{i}|\mathbf{m}\rangle_{A}|\mathbf{(N-m)}\rangle_{B}-\alpha_{i}|\mathbf{\overline{m}}\rangle_{A}|\mathbf{\overline{(N-m)}}\rangle_{B}\label{eq:m,Q,i}\end{eqnarray}
 where $\mathbf{m}$ is a $m$-bit string, and $\mathbf{(N-m)}$ is
a $(N-m)$-bit string of $\{0,1\}$; $\mathbf{\overline{m}},\mathbf{\overline{(N-m)}}$
are their bit-wise orthogonals. Corresponding to the above pair, across
the same bipartition $(m,Q)$, there also exists another \emph{unique}
conjugate pair, say, $|\psi_{j}^{\pm}\rangle$,\begin{eqnarray}
|\psi_{j}^{+}(m,Q)\rangle_{AB} & = & \alpha_{j}|\mathbf{m}\rangle_{A}|\mathbf{\overline{(N-m)}}\rangle_{B}+\beta_{j}|\mathbf{\overline{m}}\rangle_{A}|\mathbf{(N-m)}\rangle_{B}\nonumber \\
|\psi_{j}^{-}(m,Q)\rangle_{AB} & = & \beta_{j}|\mathbf{m}\rangle_{A}|\mathbf{\overline{(N-m)}}\rangle_{B}-\alpha_{j}|\mathbf{\overline{m}}\rangle_{A}|\mathbf{(N-m)}\rangle_{B}\label{eq:m,Q,j}\end{eqnarray}
 Changing to the notation $|\mathbf{m}\rangle_{A}=|\mathbf{0}\rangle_{A};|\mathbf{N-m}\rangle_{B}=|\mathbf{0}\rangle_{B};|\mathbf{\overline{m}}\rangle_{A}=|\mathbf{1}\rangle_{A};|\mathbf{\overline{N-m}}\rangle_{B}=|\mathbf{1}\rangle_{B}$
the states can be written in a compact way, \begin{eqnarray}
|\psi_{i}^{+}(m,Q)\rangle_{AB} & = & \alpha_{i}|\mathbf{0}\rangle_{A}|\mathbf{0}\rangle_{B}+\beta_{i}|\mathbf{1}\rangle_{A}|\mathbf{1}\rangle_{B}\nonumber \\
|\psi_{i}^{-}(m,Q)\rangle_{AB} & = & \beta_{i}|\mathbf{0}\rangle_{A}|\mathbf{0}\rangle_{B}-\alpha_{i}|\mathbf{1}\rangle_{A}|\mathbf{1}\rangle_{B}\nonumber \\
|\psi_{j}^{+}(m,Q)\rangle_{AB} & = & \alpha_{j}|\mathbf{0}\rangle_{A}|\mathbf{1}\rangle_{B}+\beta_{j}|\mathbf{1}\rangle_{A}|\mathbf{0}\rangle_{B}\label{eq:lemma3-eq}\\
|\psi_{j}^{-}(m,Q)\rangle_{AB} & = & \beta_{j}|\mathbf{0}\rangle_{A}|\mathbf{1}\rangle_{B}-\alpha_{j}|\mathbf{1}\rangle_{A}|\mathbf{0}\rangle_{B}\nonumber \end{eqnarray}
 It now follows from lemma 1 that the above set is completely indistinguishable,
and any subset of three states chosen from the above set is not perfectly
distinguishable. Let's emphasize that, for the above pairs of states
their indistinguishability holds strictly across the bipartition $(m,Q)$.
It can be shown that across any other bipartition, the above conjugate
pair is perfectly distinguishable.

We now show that, the entire basis can be grouped into $2^{N-2}$
unique (unique with respect to the bipartition being considered) blocks,
each block consisting of two conjugate pairs having exactly the same
LOCC distinguishability properties like those in Eq.$\,$(\ref{eq:lemma3-eq}).
To see how it's done, consider another conjugate pair $|\psi_{k}^{\pm}\rangle$,

\begin{eqnarray}
|\psi_{k}^{+}(m,Q)\rangle_{AB} & = & \alpha_{k}|\mathbf{m}\rangle_{A}|\mathbf{(N-m)'}\rangle_{B}+\beta_{k}|\mathbf{\overline{m}}\rangle_{A}|\mathbf{\overline{(N-m)'}}\rangle_{B}\nonumber \\
|\psi_{k}^{-}(m,Q)\rangle_{AB} & = & \beta_{k}|\mathbf{m}\rangle_{A}|\mathbf{(N-m)}'\rangle_{B}-\alpha_{k}|\mathbf{\overline{m}}\rangle_{A}|\mathbf{\overline{(N-m)'}}\rangle_{B}\label{eq:psi+-,k}\end{eqnarray}
 The conjugate pair $|\psi_{k}^{\pm}(m.Q)\rangle$ is different from
those in Eqs.$\;$(\ref{eq:m,Q,i}) and (\ref{eq:m,Q,j}) only in
the bit values of B. As before, there must also exist another conjugate
pair, $|\psi_{l}^{\pm}(m.Q)\rangle$, \begin{eqnarray}
|\psi_{l}^{+}(m,Q)\rangle_{AB} & = & \alpha_{l}|\mathbf{m}\rangle_{A}|\mathbf{\overline{(N-m)'}}\rangle_{B}+\beta_{l}|\mathbf{\overline{m}}\rangle_{A}|\mathbf{(N-m)'}\rangle_{B}\nonumber \\
|\psi_{l}^{-}(m,Q)\rangle_{AB} & = & \beta_{l}|\mathbf{m}\rangle_{A}|\mathbf{\overline{(N-m)'}}\rangle_{B}-\alpha_{l}|\mathbf{\overline{m}}\rangle_{A}|\mathbf{(N-m)'}\rangle_{B}\label{eq:psi+-,l}\end{eqnarray}
 Proceeding in the same way $2^{N-m-1}$ such distinct blocks can
be constructed for the same bit values of A, that is for the same
ordered pair $(\mathbf{m,\overline{m})}$. This is because, for a
fixed \emph{m-bit} string of A, there are $2^{N-m}$ possible strings
corresponding to B. Out of those $2^{N-m}$ strings, each block uses
two \emph{(N-m)-bit} strings. Therefore, the number of distinct blocks
is $2^{N-m}/2=2^{N-m-1}.$ Now there are $2^{m}$ distinct \emph{m-bit}
strings that are possible for A, however, each block uses two of them,
say $m'$ and $\overline{m'}$. Therefore, the number of distinct
blocks corresponding to different bit values of A but the same bit
values of B is $2^{m-1}$. Therefore, altogether there are $2^{m-1}.2^{N-m-1}=2^{N-2}$
distinct blocks, each block consists of two conjugate pairs and has
the same LOCC distinguishability property as the block of states specified
by Eqs. (\ref{eq:m,Q,i},\ref{eq:m,Q,j}). Now, any set consisting
of more than $2^{N-1}$ states must have at least three states from
one block. Because these three states are not perfectly distinguishable,
hence the entire set is also not perfectly distinguishable.

To show that the proof indeed holds for all bipartitions, first observe
that for another bipartition of the type $(m,Q')$ (that is a different
set of $m$ qubits are selected), only the constituent states of each
block change. The indistinguishability property of the states in each
block, which is the key feature of the entire proof, remains unaffected.
This means, the entire basis can again be grouped into $2^{N-2}$
blocks where any three states from each block are not perfectly distinguishable,
and therefore, it is not possible to pick more than $2^{N-1}$ states
and still be able to distinguish the states perfectly. For any other
bipartition with a different $m$ value, what changes are the length
of the $m-bit$ string which corresponds to the number of qubits on
Alices's side, and the constitutent states of each block; the crux
of the argument does not change. This concludes the proof. $\square$

\emph{Proof of Theorem 2.} Let's recall that a set of states is conclusively
distinguishable, if at least one state can be correctly identified
with a nonzero probability, no matter how small. Consider a set of
GHZ states containing a state whose conjugate partner is not included
in the set. This particular state can always be correctly identified
just by doing measurement in the computational basis which can indeed
by carried out even when all qubits are separated from each other.
It is easy to see that such set of states, $S$ can always be constructed
for all values of cardinality $|S|\leq2^{N}-1$. $\square$

\emph{Proof of Theorem 3.} Consider the hybrid basis consisting of
$K$ conjugate pairs, and $2^{N}-2K$ product states. Construct a
set $S$ of cardinality $|S|=2^{N}-K$ in the following way: include
all the product states, and one state from each conjugate pair. This
set is again perfectly distinguishable by LOCC when all qubits are
spatially separated, and therefore for all spatial configurations.
To show that any set of cardinality greater than $2^{N}-K$ is not
perfectly distinguishable, first note that a conjugate pair $|\psi_{k}^{\pm}\rangle$
is reduced to a product pair by setting $\beta_{k}=0$. Recall that
when the entire basis was entangled, then across every bipartition
we were able to group the states into $2^{N-2}$ blocks, each block
containing two conjugate pairs. For the hybrid case, some of the blocks
now contain either one conjugate pairs and two product states (the
same product states whose superposition would have given rise to the
corresponding conjugate pair in the all-entangled case) or four product
states (these product states are those corresponding to the two conjugate
pairs in the all-entangled case).

If one more state is added, then the set now includes either a block
of four states containing one conjugate pair and two product states
(because all product states, and one state from each conjugate pair
have already been included) which is not perfectly distinguishable
(by lemma 2), or a block of three states containing one conjugate
pair and one state from another conjugate pair which are not perfectly
distinguishable (by lemma 1). Note that we just decribed the worst
case scenario, and other sets containing $2^{N}-K$ states would invariably
contain a block (s) which is(are) not perfectly distinguishable either
by lemma 1 or lemma 2. This holds for all bipartitions (see the proof
of Theorem 1), and therefore for all spatial configurations. $\square$

To conclude, it is shown that for GHZ states, entanglement content
is not a contributing factor in determining the maximum number of
states that are perfectly distinguishable by LOCC, although, entanglement
certainly is. This goes against our intution and inequality (\ref{Hayashi-II}),
in the sense both suggest a larger number of weakly entangled states
can be perfectly distinguished than strongly entangled ones. One surprising
feature, in the context of the result obtained is the spatial configuration
independence of the threshold value. We have also shown a hybrid ensemble
comprising both entangled GHZ states and the product states is less
indistinguishable in the sense, the upper bound on the maximum number
of perfectly distinguishable states is always greater than $2^{N-1}$.
In particular the upper bound approaches the total dimension as more
and more product states are included.

The open question is whether entanglement independence of the upper
bound is a generic feature. Supporting evidence might be obtained
by looking into the non-maximal canonical basis in $d\otimes d$.
This basis, where every state has Schmidt rank $d$ is a direct generalization
of the $2\otimes2$ non-maximal canonical basis considered in lemma
1. Taking cue from the distinguishability of maximally entangled canonical
basis in $d\otimes d$ \cite{Ghosh2002} and the result presented
here, we could expect that no more than $d$ states can be perfectly
distinguished irrespective of the average entanglement. However, if
we mix entangled states of different Schmidt ranks the upper bound
is likely to increase, but the exact functional form is not immediately
clear. 
\begin{acknowledgments}
Many thanks to Guruprasad Kar, and Jonathan Walgate for useful discussions,
and to Shashank Virmani for his comments on an earlier version of
this work. This research is supported by Canada's Natural Sciences
and Engineering Research Council (NSERC). \end{acknowledgments}

\end{document}